\documentclass[11pt]{article}
\usepackage{amsmath, amssymb}
\usepackage{enumerate}
\numberwithin{equation}{section}
\allowdisplaybreaks[1]

\usepackage{graphicx}
\newcommand{\p}{\partial}

\newcommand{\bt}{{\bf t}}

\newcommand{\NN}{\mathbb{N}}

\newcommand{\CC}{\mathbb{C}}
\newcommand{\ZZ}{\mathbb{Z}}

\newcommand{\A}{\mathcal{A}}

\newcommand{\beq}{\begin{equation}}
\newcommand{\eeq}{\end{equation}}

\newcommand{\nn}{\nonumber}

\newtheorem{dfn}{Definition}[section]
\newtheorem{lem}[dfn]{Lemma}
\newtheorem{prp}[dfn]{Proposition}
\newtheorem{thm}[dfn]{Theorem}
\newtheorem{rmk}[dfn]{Remark}

\newtheorem{emp}[dfn]{Example}

\newenvironment{prf}{\noindent {\it Proof} \ }{\hfill $\Box$}
\newenvironment{prfn}[1]{\noindent {\it Proof of #1} \ }{\hfill $\Box$}

\DeclareMathOperator{\res}{{\rm res}\,}

\begin{document}

\title{Gelfand--Dickey hierarchy,  generalized BGW \\ tau-function, and $W$-constraints}
\author{Di Yang, \quad Chunhui Zhou}
\date{}\maketitle
\begin{abstract}
Let $r\geq 2$ be an integer. The generalized BGW 
tau-function for the Gelfand--Dickey hierarchy of $(r-1)$ dependent variables (aka the $r$-reduced KP hierarchy) is defined as 
a particular tau-function 
that depends on $(r-1)$ constant parameters $d_1,\dots,d_{r-1}$.
In this paper we show that this tau-function satisfies a family of linear equations, 
called the $W$-constraints of the second kind. The operators giving rise to the linear equations 
also depend on~$(r-1)$ constant parameters. We show that there is a one-to-one correspondence 
between the two sets of parameters. 
\end{abstract} 

\section{Introduction}
Let $r\geq2$ be an integer, and let $n=r-1$. The 
Gelfand--Dickey (GD) hierarchy with $n$ unknown functions is an infinite family of PDEs, defined by  
\beq\label{GD}
 \frac{\p L}{\p t_i}=\biggl[\Bigl(L^{i/r}\Bigr)_+,L\biggr],\quad 
 i\in \NN\backslash r\NN,
\eeq
where 
\beq
L := \p^{r}+v_1\p^{r-2}+\cdots+v_{r-1}
\eeq 
is the Lax operator, $L^{i/r}$, $i\in \NN\backslash r\NN$, denote the fractional 
powers of~$L$ (cf.~e.g.~\cite{Dickey} for the definition), and $\p$ is understood as $\p_{t_1}$. 
This integrable hierarchy can also be viewed as a reduction of the Kadomtsev--Petviashvili (KP) hierarchy (see e.g.~\cite{Dickey} or Section~\ref{section2}). 

There are many interesting solutions to the GD hierarchy. For example, 
in the study of Witten's $r$-spin invariants~\cite{FSZ, G, W2}, 
the so-called {\it topological solution}~\cite{Du,DZ-norm} to the GD hierarchy plays an important role.
For example, for the case $r=2$, the GD hierarchy is the celebrated Korteweg--de Vries (KdV) hierarchy, and the topological solution 
is famously known as the Witten--Kontsevich solution (cf.~e.g.~\cite{DYZ, DZ-norm}), governing the integrals of 
psi-classes over the moduli space of curves~\cite{K, W1}. The interest of this 
paper is on another solution to the GD hierarchy again for any~$r$. Unlike 
the topological solution, the 
solution of interest of this paper will depend non-trivially on $r-1$ arbitrary parameters. 
Again, let us look at the KdV case first (i.e., the case with $r=2$). For this case, 
it is known that there exists a solution to the KdV hierarchy, 
called the generalized BGW solution, depending non-trivially on one arbitrary parameter~\cite{A,DYZ},  
having bispectral properties~\cite{DYZ, DG}, 
and possessing enumerative meanings 
\cite{KN, N, YZ21}. This motivates us to generalize the 
generalized BGW solution to an arbitrary $r\geq2$. Indeed, let $d_1,\dots,d_n$ be arbitrarily given complex numbers, and define 
 $v_{\rm BGW}(\bt)$ as the unique solution in $\CC[[\bt]]^{n}$ to the GD hierarchy,  
satisfying the initial condition
\begin{align}
\label{initial-condition} & 
v_{\alpha,{\rm BGW}}\bigl(t_1,\bt_{\geq2}={\bf 0}\bigr)=\frac{d_\alpha}{(1-t_1)^{\alpha+1}},\quad \alpha=1,\dots,r-1,
\end{align}
where $\bt = (t_i)_{i\in\NN\backslash r\NN}$. We call $v_{\rm BGW}(\bt;d_1,\dots,d_n)$ the {\it generalized 
BGW solution to the GD hierarchy}. The Dubrovin--Zhang type 
tau-function of this solution (cf.~\cite{BDY, BY, Dickey, DYZ, DZ-norm}) will be called the 
{\it generalized BGW tau-function}, denoted by $\tau_{\rm BGW}=\tau_{\rm BGW}(\bt; d_1,\dots,d_n)$. 
We show in Section~\ref{section3} that the generalized BGW tau-function $\tau_{\rm BGW}$ can be chosen such that 
\beq\label{new-string}
\sum_{i\in\NN\backslash r\NN}i \tilde t_ i\frac{\p\tau_{\rm BGW}}{\p t_i}+\frac{d_1}{r} \tau_{\rm BGW}=0,
\eeq
where~$ \tilde t_i=t_i-\delta_{i,1}$, moreover, it is unique up to multiplying by a nonzero constant. 

The above definition of the generalized BGW tau-function for the GD hierarchy was given in joint work 
by B.~Dubrovin, D.~Zagier and the first named author of the present paper in a more general 
set up, i.e., for the Drinfeld--Sokolov (DS) hierarchy associated to a simple Lie algebra~\cite{BDY}, where 
certain analogues of the triangle numbers on the constants manifold were observed. 
The GD hierarchy can be considered as the DS hierarchy associated to the $A_n$ type simple 
Lie algebra under the Wronskian gauge. 
In~\cite{LWZ}, certain generalized BGW tau-functions were also given for the DS hierarchy associated 
to an affine Kac--Moody algebra (the simple Lie algebra case corresponds to the untwisted 
affine Kac--Moody algebra under a particular choice of vertex in the Dynkin diagram).

For $r=2$, 
the generalized BGW tau-function can also be identified with the solution to  
Virasoro constraints~\cite{A, BR, BBCCN}. 
The goal of this paper is to show that for an arbitrary $r\geq2$, the 
generalized BGW tau-function (defined above) satisfies a set of linear constraints, which will be called 
{\it $W$-constraints of the second kind}. 
To be precise, define a family of operators $W^{\rm red}_{\alpha,q}$, $\alpha=1,\dots,n=r-1$, $q\geq 0$, by 
\beq\label{defJ}
W^{\rm red}_{\alpha,q}({\bf t}):=
\res_{\lambda}\lambda^{\alpha+ (q-\alpha) r}\Bigl(\p_{\mu}^{\alpha+1} \bigl( X_{\rm GD}(\tilde {\bf t};\lambda,\mu)\bigr)\Bigr)\Big|_{\mu=\lambda} d\lambda,
\eeq
where $X_{\rm GD}({\bf t};\lambda,\mu)$ is given by
\beq\label{vertex-GD}
X_{\rm GD}({\bf t};\lambda,\mu):=
e^{\sum_{i\in\NN\backslash r\NN}  t_i\left(\mu^i-\lambda^i\right)} \circ
e^{\sum_{i\in\NN\backslash r\NN}\left(\frac1{i\lambda^i}-\frac1{i\mu^i}\right)\frac{\p}{\p t_i}}. 
\eeq
These operators were given e.g.~in~\cite{AM}; according to~\cite{AM,BM0,FKN}, they can be expressed by 
operators coming from the twisted module of the $\mathcal{W}_{A_n}$-algebra~\cite{BM}.
We have the following theorem.

\begin{thm}\label{thmwconst}
There exist unique constants $\rho_1,\dots,\rho_{r-1}\in\CC$ such that
\beq\label{w-eq-1}
W^{\rm red}_{\alpha,q} \bigl(\tau_{\rm BGW}\bigr)=(-1)^\alpha \rho_\alpha \delta_{\alpha,q}\tau_{\rm BGW}, \qquad \alpha=1,\dots,r-1,~ q\geq \alpha.
\eeq
Moreover, these constants $\rho_\alpha$ are polynomials of $d_1,\dots,d_n$, having the form
\beq\label{prprho}
\rho_\alpha=\frac{d_\alpha}r+\omega_\alpha(d_1,\dots,d_{\alpha-1}).
\eeq
\end{thm}

We refer to~\eqref{w-eq-1} as {\it $W$-constraints of the second kind}. 
We note that the $W$-constraints for the topological tau-function~\cite{AM, BM0, BM, BBCCN, Goe, Zhou}, being 
referred to as $W$-constraints of the first kind, start with $q=0$ instead of $q=\alpha$ 
and have the dilaton shift at $t_{r-1}$ instead of at~$t_1$.

The $W$-constraints of the second kind seem to have 
common cases with the $W$-constraints given in~\cite{BBCCN} in an equivalent way.
For certain special common cases,  
the solutions to the $W$-constraints of~\cite{BBCCN} were conjectured 
by Chidambaram, Garcia--Failde and Giacchetto in a recent letter to the authors of the present paper 
to be tau-functions for the GD hierarchy.  
Theorem~\ref{thmwconst} (cf.~also Theorem~\ref{thm-main}) should lead to this conjecture; still, it will be 
interesting to investigate the explicit relationship between $\tau_{\rm BGW}$ and the partition functions defined in~\cite{BBCCN}.

In a subsequent publication, we will consider the analogous open extension of the generalized BGW tau-function for 
arbitrary $r\geq2$ (see~\cite{YZ} for the $r=2$ case; cf.~also~\cite{BY, Buryak}). 

\paragraph{Organization of the paper.} 
In Section~\ref{section2} we review some basics on KP and GD hierarchies. In Section~\ref{section3} 
we give the definition of the generalized BGW tau-function $\tau_{\rm BGW}$ in more details. In Section~\ref{section4} 
we prove Theorem~\ref{thmwconst}. In Section~\ref{section5} we present some examples. 

\paragraph{Acknowledgements.}
We thank Ga\"etan Borot and Xingjun Lin for helpful suggestions. The work is partially supported by 
 NSFC 12061131014.

\section{Preliminaries}\label{section2}
In this section we review 
tau-functions and wave functions for the KP hierarchy 
and for the GD hierarchy.

Let $L_{\rm KP}$ denote the pseudo-differential operator 
\beq\label{Lax-operator}
L_{\rm KP}:=\p+\sum_{k\geq1}u_k \p^{-k}.
\eeq
Here $\p:=\p_x$. 
Recall that the KP hierarchy~\cite{Dickey} is the following 
commuting system of PDEs for the 
infinitely many dependent variables 
$u_1({\bf t}_{\rm KP}),u_2({\bf t}_{\rm KP})$, $\dots$: 
\beq\label{KP}
\frac{\p L_{\rm KP}}{\p t_i}=\Bigl[\bigl(L_{\rm KP}^i\bigr)_+,L_{\rm KP}\Bigr],\quad i\geq1.
\eeq
Here ${\bf t}_{\rm KP}:=(t_1,t_2,t_3,\dots)$ denotes the infinite vector of times.
The first equation in~\eqref{KP} reads
\[ \frac{\p u_k}{\p t_1}=\frac{\p u_k}{\p x},\quad k\geq1.\] 
Therefore we identify the time~$t_1$ with~$x$.
We consider solutions to the KP hierarchy in $\CC[[{\bf t}_{\rm KP}]]^{\NN}$, i.e., 
$u_k({\bf t}_{\rm KP})\in \CC[[{\bf t}_{\rm KP}]]$, $k\geq 1$. Denote for simplicity 
 $u:=(u_1,u_2,\cdots)$.   
It is known (see for example~\cite{Dickey}) that 
for an arbitrary power series solution 
$u({\bf t}_{\rm KP})=\left(u_1({\bf t}_{\rm KP}),u_2({\bf t}_{\rm KP}),\dots\right)$ 
to the KP hierarchy, 
there exists a pseudo-differential operator
\beq
\Phi({\bf t}_{\rm KP})=1+\sum_{k\geq1}\phi_k({\bf t}_{\rm KP}) \, \p^{-k},
\quad \phi_k({\bf t}_{\rm KP}) \in\CC[[{\bf t}_{\rm KP}]],
\eeq
called a dressing operator, satisfying 
\begin{align}
& L_{\rm KP}=\Phi \circ \p \circ \Phi^{-1}, \\ 
& \frac{\p \Phi}{\p t_i}= - \bigl(L_{\rm KP}^i\bigr)_- \circ \Phi,\quad i\geq1.
\end{align}
The dressing operator~$\Phi$ is uniquely determined by the solution $u$ up 
to the right multiplication by an operator of the form
\[
1+\sum_{k\geq1}a_k\p^{-k}\in\CC[[\p^{-1}]],
\]
where $a_k$, $k\geq 1$ are constants. 
The wave and dual wave functions $\psi({\bf t}_{\rm KP};\lambda)$, $\psi^*({\bf t}_{\rm KP};\lambda)$ associated to the solution~$u({\bf t}_{\rm KP})$
are elements in $\CC\bigl(\bigl(\lambda^{-1}\bigr)\bigr)[[{\bf t}_{\rm KP}]]$ defined by 
\begin{align}
& \psi({\bf t}_{\rm KP};\lambda):=\Phi({\bf t}_{\rm KP};\lambda) \Bigl(e^{\xi({\bf t}_{\rm KP};\lambda)}\Bigr), \quad
 \psi^*({\bf t}_{\rm KP};\lambda)=\bigl(\Phi^*({\bf t}_{\rm KP};\lambda)\bigr)^{-1} \Bigl(e^{-\xi({\bf t}_{\rm KP};\lambda)}\Bigr), \label{wave-dual-wave-fct}
\end{align}
where $\xi({\bf t}_{\rm KP};\lambda):=\sum_{i\geq1}t_i\lambda^i$, 
and $\Phi^*$ denotes the formal adjoint operator 
of~$\Phi$, i.e.,
\beq
\Phi^*:=1+\sum_{k\geq1}(-\p)^{-k}\circ \phi_k.
\eeq
They satisfy 
\begin{align*}
&L_{\rm KP}(\psi)=\lambda \psi,\quad 
\frac{\p  \psi}{\p t_i}=\left(L_{\rm KP}^i\right)_+(\psi),\\
&L_{\rm KP}^*(\psi^*)=\lambda \psi^*,\quad
\frac{\p\psi^*}{\p t_i}=\Bigl(\bigl(L_{\rm KP}^*\bigr)^i\Bigr)_+(\psi^*)
\end{align*}
with $L_{\rm KP}^*:=-\p+\sum_{k\geq1}(-\p)^k\circ u_k$. 
Introduce the following two operators:
\beq
X({\bf t}_{\rm KP};\lambda)
=e^{\sum_{i\geq1}t_i\lambda^i}e^{-\sum_{i\geq1}\frac{1}{i\lambda^i}\p_i},\quad
X^*({\bf t}_{\rm KP};\lambda)
=e^{-\sum_{i\geq1}t_i\lambda^i}e^{\sum_{i\geq1}\frac{1}{i\lambda^i}\p_i}.
\eeq 

It was proved in~\cite{Dickey} that 
for an arbitrary solution $u$ in $\CC[[{\bf t}_{\rm KP}]]^{\NN}$ to the KP hierarchy, 
there exists a power series $\tau_{\rm KP}({\bf t}_{\rm KP})\in\CC[[{\bf t}_{\rm KP}]]$, satisfying 
\beq
\psi({\bf t}_{\rm KP};\lambda) =\frac{X({\bf t}_{\rm KP};\lambda)(\tau_{\rm KP}({\bf t}_{\rm KP}))}{\tau_{\rm KP}({\bf t}_{\rm KP})},\quad
\psi^*({\bf t}_{\rm KP};\lambda)=\frac{X^*({\bf t}_{\rm KP};\lambda)(\tau_{\rm KP}({\bf t}_{\rm KP}))}{\tau_{\rm KP}({\bf t}_{\rm KP})}.
\eeq
We call $\tau_{\rm KP}({\bf t}_{\rm KP})$ the tau-function of the solution~$u$ for the KP hierarchy. 
We also call $(\Phi,\tau_{\rm KP})$ a dressing pair associated to~$u$.
The dressing pair is uniquely determined by the solution~$u$ up to the transformation
\[
\left(\Phi,\tau_{\rm KP}\right)\mapsto
\left(\Phi \circ e^{-\sum_{i\geq1} b_i\p^{-i}},\quad \tau_{\rm KP} e^{b_0+\sum_{i\geq1} b_i t_i}\right), \quad b_0,b_1,b_2,\dots\in\CC.
\]

Denote by $\A_u:=\A_{u,0}\bigl[\p^i (u_k)| i,k\geq1\bigr]$ 
the ring of differential polynomials of~$u$, 
where $\A_{u,0}$ denotes the ring of smooth functions of~$u$. 
For a pseudo-differential operator of the form $a=\sum_{i\in\ZZ}a_i\p^i$, 
define $\res_{\p}a=a_{-1}$. 
Define a family of differential polynomials in $u$ by~\cite{Dickey}
\beq\label{def-omega}
\Omega_{i,j}^{\rm KP}=\Omega_{i,j}^{\rm KP}(u,u_x,\dots):=\p^{-1}\biggl(\frac{\p}{\p t_j}\res_{\p} L_{\rm KP}^i\biggr) \in \mathcal{A}_u,
\quad i,j\geq1,
\eeq
where 
$\p^{-1}$ is fixed by the no-integration-constant rule. We call 
$\Omega_{i,j}$ the two-point correlations functions for the KP hierarchy.  

\begin{lem}\label{lem-tau-omega}
Let $u$ be an arbitrary solution in $\CC[[{\bf t}_{\rm KP}]]^{\NN}$ to the KP hierarchy, 
and $\tau_{\rm KP}\in\CC[[{\bf t}_{\rm KP}]]$ the tau-function of~$u$. 
Then the following formulae hold true:
\beq
\frac{\p^2\log\tau_{\rm KP}}{\p t_i \p t_j} = \Omega_{i,j}^{\rm KP}, \quad \forall\,i,j\geq1. 
\eeq
\end{lem}
\begin{prf} 
Let $(\Phi,\tau_{\rm KP})$ be the dressing pair associated to~$u$, and 
$\psi$ the corresponding wave function. 
It was shown in~\cite{Dickey} that for given $i,j\geq1$, 
\beq\label{omega-wave}
\frac{\p^2 \log\tau_{\rm KP}}{\p t_i \p t_j}
	=\res_z z^i\biggl(-\sum_{\ell\geq1}z^{-\ell-1}\p_{t_\ell}+\frac{\p}{\p z}\biggr) 
	\Biggl(-\frac{\big(L_{\rm KP}^j\big)_-(\psi)}{\psi}\Biggr)
\eeq
Note that $\bigl(L_{\rm KP}^j\bigr)_-$ can be rewritten into the form
\beq \label{LiminusequalLminus}
	\bigl(L_{\rm KP}^j\bigr)_-=\sum_{k\geq1}a_{j,k} L_{\rm KP}^{-k},
\eeq
where $a_{j,k}\in\A_u$ satisfying $a_{j,k}|_{u=u_x=u_{xx}=\dots=0}=0$.
Combining~\eqref{omega-wave} with~\eqref{LiminusequalLminus} we find 
that 
\[
\frac{\p^2 \log\tau_{\rm KP}}{\p t_i \p t_j}=i a_{j,i}+\sum_{k=1}^{i-1}\frac{\p a_{j,k}}{\p t_{i-k}}.
\]
In particular, observing that $a_{j,1}=\res_{\p} L_{\rm KP}^j$, we have
\[
\frac{\p^2 \log\tau}{\p t_1 \p t_j}=\res_{\p}L_{\rm KP}^j.
\]
Taking the derivative with respect to $t_i$ on the both sides of the above identity, 
and then by using the definition~\eqref{def-omega}, 
the lemma is proved. 
\end{prf}

Introduce the following operator:
\beq\label{def-X}
X\left({\bf t}_{\rm KP};\lambda,\mu\right):=
e^{\sum_{i\geq1}t_i\left(\mu^i-\lambda^i\right)} \circ
e^{\sum_{i\geq1}\left(\frac{1}{i\lambda^i}-\frac{1}{i\mu^i}\right)\frac{\p}{\p t_i}}.
\eeq
It is shown in~\cite{DJKM} that
\begin{align}
&\res_\nu \frac{X({\bf t}_{\rm KP};\nu)\circ X({\bf t}_{\rm KP};\lambda,\mu)\left(\tau_{\rm KP}({\bf t}_{\rm KP})\right)}{\tau_{\rm KP}({\bf t}_{\rm KP})} 
\frac{X^*({\bf t}_{\rm KP}';\nu)\left(\tau_{\rm KP}({\bf t}_{\rm KP}')\right)}{\tau_{\rm KP}({\bf t}_{\rm KP}')} d\nu
\nn\\
 =&(\lambda-\mu)\psi({\bf t}_{\rm KP};\mu)\psi^*({\bf t}_{\rm KP}';\lambda),
\label{double-wave}
\end{align}
where ${\bf t}_{\rm KP}=(t_1,t_2,\dots)$ and ${\bf t}_{\rm KP}'=(t_1',t_2',\dots)$.

Following~\cite{OS} and~\cite{ASM}, introduce the following operator: 
\beq\label{OS}
M:=\Phi \circ \biggl(\sum_{i\geq1}i t_i \p^{i-1}\biggr)\circ\Phi^{-1}\in \CC[[{\bf t}_{\rm KP}]]\otimes_{\CC} \CC((\p^{-1})).
\eeq
By using Lemma~3.2 of~\cite{ASM} we have
\beq\label{residue-ml}
\res_{\p}M^i \circ L_{\rm KP}^k
=\res_{\lambda} \lambda^k
\psi^*({\bf t}_{\rm KP};\lambda)
\p_{\lambda}^i (\psi({\bf t}_{\rm KP}; \lambda)),\quad \forall~i,k\geq0. 
\eeq
Then it follows from~\eqref{double-wave} that 
\beq\label{residue-ml0}
\res_{\p}M^i \circ L_{\rm KP}^k
=\frac{1}{i+1}\p
\left(
\frac{
\res_{\lambda} \lambda^k \p_{\mu}^{i+1}\circ X({\bf t}_{\rm KP};\lambda,\mu)\left(\tau_{\rm KP}({\bf t}_{\rm KP})\right)\mid_{\mu=\lambda}}
{\tau_{\rm KP}({\bf t}_{\rm KP})}
\right).
\eeq

Denote by ${\bf t}=\left(t_i\right)_{i\in\NN\backslash r\NN}$ 
the infinite time vector for the GD hierarchy. 

For an arbitrary solution 
$v({\bf t})
=\bigl(v_1({\bf t}),\dots,v_{r-1}({\bf t})\bigr)$
in $\CC[[{\bf t}]]^{r-1}$ to the GD hierarchy, 
we associate to it an infinite sequence of power series in $\CC[[{\bf t}]]$ defined by 
\[
u_k=u_k({\bf t}):=\res_{\p} \Bigl(L^{1/r}\circ\p^{k-1}\Bigr),\quad k\geq1.
\]
In other words,
\beq\label{KP-GD}
\p+\sum_{k\geq1}u_k\p^{-k}=L^{1/r}. 
\eeq
Obviously, $u=(u_1,u_2,\dots)$ satisfies the KP hierarchy, namely,  
for all $i\in \NN$,
\beq\label{gdh}
\frac{\p L^{1/r}}{\p t_i}=\biggl[\Bigl(L^{i/r}\Bigr)_+,\,L^{1/r}\biggr].
\eeq
Let $\tau_{\rm KP}$ be the tau-function of the solution $u$ 
to the {\it KP hierarchy}.  
By the definition~\eqref{def-omega}, we know $\Omega_{ir,jr}^{\rm KP}=0$ for $i,j\geq1$ when $L_{\rm KP}=L^{1/r}$. 
It then follows from Lemma~\ref{lem-tau-omega} that $\tau_{\rm KP}$ satisfies
\[
\frac{\p^2 \log\tau_{\rm KP}}{\p t_{i r} \p t_{j r}}=0,\quad i,j\geq1.
\]
This means that, there exist constants $a_1,a_2,\dots$, such that
\[
\frac{\p}{\p t_{ir}}\left(\sum_{j\geq1}a_j t_{j r}+\log\tau_{\rm KP}\right)=0, \quad i\geq1.
\]
Let $\tau=\tau({\bf t}):=\tau_{\rm KP}({\bf t}_{\rm KP}) \exp\left(\sum_{j\geq1}a_j t_{jr}\right)$. 
Then $\tau$ is still a KP tau function. 
We call this particular chosen~$\tau$ {\it the tau-function of the solution~$v$ for the GD hierarchy reduced from the KP hierarchy}.

Let us proceed to give a second definition of tau-function for the GD hierarchy.
\begin{lem} \label{lemtaudz}
For an arbitrary solution $v=v(\bt)$ in~$\CC[[{\bf t}]]^{r-1}$ to the GD hierarchy~\eqref{GD}, 
there exists a power series $\tau_{\rm DZ}=\tau_{\rm DZ}(\bt)\in \CC[[\bt]]$ satisfying 
\beq\label{def-tau-DZ}
\frac{\p^2\log \tau_{\rm DZ}}{\p t_i \p t_j}= \Omega_{i,j}^{\rm GD},\qquad \forall\, i,j\in \NN\backslash r\NN,
\eeq
where $\Omega_{i,j}^{\rm GD}$ are differential polynomials in $v$ defined as
\beq\label{def-omega-GD}
\Omega^{\rm GD}_{i,j}=\Omega^{\rm GD}_{i,j}(v,v_x,\dots):=\p^{-1}\left(\frac{\p \res_{\p}L^{i/r}}{\p t_j}\right).
\eeq
Here $\p^{-1}$ is again fixed by the no-integration-constant rule.
We call $\tau_{\rm DZ}$ the Dubrovin--Zhang type tau-function of the solution $v$ 
for the GD hierarchy. 
\end{lem}
Note that $\tau_{\rm DZ}$ is uniquely determined by~$v$ 
up to multiplying by a factor the form 
\beq\label{formexp}
\exp\biggl(b_0 + \sum_{i \in \NN\backslash r\NN} b_i t_i\biggr) \,, 
\eeq
where $b$'s are arbitrary constants. 

\begin{thm}\label{equiv-red-DZ}
Let $v$ be an arbitrary solution in $\CC[[\bt]]^{r-1}$ to the GD hierarchy, 
and $\tau_{\rm DZ}$ and $\tau$ be 
the Dubrovin--Zhang type tau-function 
and the tau-function reduced from the KP hierarchy of~$v$, respectively. 
Then there exist constants $b_0,b_1,b_2,\dots$ such that 
\beq
\tau = \tau_{\rm DZ} \exp\biggl(b_0 + \sum_{i \in \NN\backslash r\NN} b_i t_i\biggr).
\eeq
\end{thm}
\begin{prf}
It suffices to show $\tau$ satisfies~\eqref{def-tau-DZ}. 
Let $u$ be the solution to the KP hierarchy determined by $v$ via~\eqref{KP-GD}. 
By the definition of $\Omega_{i,j}^{\rm KP}$ and $\Omega_{i,j}^{\rm GD}$, we have
\[
\Omega_{i,j}^{\rm KP}(u,u_x,\dots)=\Omega_{i,j}^{\rm GD}(v,v_x,\dots),\quad
\forall~i,j\in\NN\backslash r\NN. 
\]
On the other hand, by using Lemma~\ref{lem-tau-omega} and using the definition of~$\tau$, 
we know $\tau$ satisfies
\[
\frac{\p^2 \log\tau}{\p t_i \p t_j}=\Omega_{i,j}^{\rm KP}(u,u_x,\dots), \quad 
\forall~i,j\in\NN\backslash r\NN,
\]
where $\Omega_{i,j}^{\rm KP}$ is given by~\eqref{def-omega}.
\end{prf}

\section{The generalized BGW tau-function for the GD hierarchy}\label{section3}
In this section we give more details about the definition of the generalized BGW tau-function.

Introduce a gradation on $\A_v$ by assigning the degree:
\[
 \deg\, \p^k (v_\alpha)=\alpha+1+k,\qquad \alpha=1,\dots,r-1,\quad k\geq0.
\]
It is easy to verify that
\beq\label{degree-res}
 \deg\, \res_{\p}\left(L^{i/r}\p^{-k}\right)=i-k,\qquad
i\geq1,\quad k\leq i-2.
\eeq
This implies the GD hierarchy~\eqref{GD} has the form
\[
\frac{\p v_\alpha}{\p t_i}=X_\alpha^i(v,v_x,\dots),
\qquad \alpha=1,\dots,r-1, \quad i\in\NN\backslash r\NN,
\]
with $X^i_\alpha=X_\alpha^i(v,v_x,\dots)\in\A_v$ having the degree
\beq\label{degree-X}
 \deg X^i_\alpha=\alpha+i+1.
\eeq

\begin{prp}\label{prp-bgw-solution}
The generalized BGW solution $v_{\rm BGW}$ satisfies the following linear equations:
\beq
\sum_{i\in\NN\backslash r\NN}i \tilde t_i\frac{\p v_{\alpha,\rm BGW}}{\p t_i}+(\alpha+1) v_{\alpha,\rm BGW}=0.
\eeq
\end{prp}
\begin{prf}
For simplicity, we denote $v_\alpha=v_{\alpha,\rm BGW}$, and denote 
\beq\label{string-test}
f_\alpha({\bf t}):=\sum_{i\in\NN\backslash r\NN}i \tilde t_i \frac{\p v_\alpha({\bf t})}{\p t_i}+(\alpha+1)v_\alpha({\bf t}).
\eeq
We are to show $f_\alpha=0$.
Firstly, from the initial condition~\eqref{initial-condition},
it is easy to see that  
\beq\label{string-test-1}
f_\alpha(t_1=x,t_2=0,\dots)=(x-1)\p_x (v_\alpha(x,{\bf 0}))+(\alpha+1)v_\alpha(x,{\bf 0})=0.
\eeq
Taking the derivative of~\eqref{string-test} 
with respect to $t_j$, 
we have
\begin{align*}
\frac{\p f_\alpha}{\p t_j}=&\sum_{i\in\NN\backslash r\NN} i\tilde t_i\frac{\p^2 v_\alpha}{\p t_i \p t_j} +(\alpha+j+1)\frac{\p v_\alpha}{\p t_j}
\\
=&\sum_{i\in\NN\backslash r\NN} i\tilde t_i 
\sum_{\beta=1}^{r-1} \sum_{k\geq0} \p^k \big(X_\beta^i\big) \frac{\p X_\alpha^j}{\p v_\beta^{(k)}}
+(\alpha+j+1)\frac{\p v_\alpha}{\p t_j}
\\
=&\sum_{\beta=1}^{r-1}\sum_{k\geq0}
\left(\p_x^k (f_\beta)-(\beta+k+1)v_\beta^{(k)}\right)
\frac{\p X^j_\alpha}{\p v_\beta^{(k)}}
+(\alpha+1+j)X^j_\alpha\\
=&\sum_{\beta=1}^{r-1}\sum_{k\geq0}\p_x^k (f_\beta)\frac{\p X^j_\alpha}{\p v_\beta^{(k)}}.
\end{align*}
Here, $v_\beta^{(k)}=\p^k(v_\beta)$, $k\geq0$, the second equality is due to
\[
\frac{\p}{\p t_i}=\sum_{\beta=1}^{r-1}\sum_{k\geq0} \p^k \big(X_\beta^i\big)
\frac{\p}{\p v_\beta^{(k)}},
\]
the third equality can be obtained by applying $\p^k$ to~\eqref{string-test},
and the last equality is due to~\eqref{degree-X}. 
Hence the identity~\eqref{string-test-1} implies
\[
\frac{\p f_\alpha}{\p t_j}(x,{\bf 0})=0. 
\] 
	
By induction on $m$, we have for arbitrary $j_1,\dots,j_m\in\NN\backslash r\NN$, 
\[
\frac{\p^m f_\alpha}{\p t_{j_1}\cdots\p t_{j_m}}(x,{\bf 0})=0.
\]
The proposition is proved. 
\end{prf}

\begin{thm}\label{exis-bgw-tau}
The generalized BGW tau-function $\tau_{\rm BGW}$ can be chosen to satisfy~\eqref{new-string}.
\end{thm}

\begin{prf}
By using~\eqref{def-omega-GD} and~\eqref{degree-res}, one can verify that
\[
\deg \Omega_{j_1,j_2}^{\rm GD}=j_1+j_2,\quad j_1,j_2\in\NN\backslash r\NN.
\]
Then it follows from Proposition~\ref{prp-bgw-solution} that
\begin{align*}
\sum_{i\in\NN\backslash r\NN} i \tilde t_i \frac{\p \Omega^{\rm GD}_{j_1,j_2}}{\p t_i}
=&\sum_{i\in\NN\backslash r\NN} i \tilde t_i \sum_{\alpha=1}^{r-1}\sum_{k\geq1}
\p_{t_i} \big(v_\alpha^{(k)}\big) \frac{\p \Omega^{\rm GD}_{j_1,j_2}}{\p v_\alpha^{(k)}}\\
=&-\sum_{\alpha=1}^{r-1}\sum_{k\geq1}(\alpha+k+1)v_\alpha^{(k)}\frac{\p \Omega^{\rm GD}_{j_1,j_2}}{\p v_\alpha^{(k)}}
=-(j_1+j_2)\Omega^{\rm GD}_{j_1,j_2}. 
\end{align*}
Therefore, by using~\eqref{def-tau-DZ}, we have
\[
\frac{\p^2}{\p t_{j_1} \p t_{j_2}} 
\Biggl(\sum_{i\in\NN\backslash r\NN}i \tilde t_i \frac{\p \log\tau_{\rm BGW}}{\p t_i} \Biggr)=0,
\quad j_1,j_2\in\NN\backslash r\NN.
\]
Hence there exist constants $a_0$ and $a_i$, $i\in\NN\backslash r\NN$, such that
\[
\sum_{i\in \NN\backslash r\NN} i \tilde t_i \frac{\p \log \tau_{\rm BGW}}{\p t_i}
=\sum_{i\in \NN\backslash r\NN} a_i \tilde t_i+a_0.
\]
Let us modify $\tau_{\rm BGW}$ as $\tau_{\rm BGW} \exp\big(-\sum_{i\in\NN\backslash r\NN}a_i t_i\big)$. 
Then $\tau_{\rm BGW}$ is still a Dubrovin--Zhang type tau-function,
and satisfies
\beq\label{determine-const}
\sum_{i\in \NN\backslash r\NN} i \tilde t_i \frac{\p \log \tau_{\rm BGW}}{\p t_i}-a_0=0.
\eeq 
	
It remains to show $a_0=-d_1/r$. 
The above~\eqref{determine-const} implies
\[
\left.\frac{\p^2 \log\tau_{\rm BGW}}{\p t_1^2}\right|_{{\bf t}={\bf 0}}
=\left.\frac{\p \log\tau_{\rm BGW}}{\p t_1}\right|_{{\bf t}={\bf 0}}
=-a_0.
\]
Hence by using~\eqref{def-tau-DZ}, \eqref{def-omega-GD}
and the initial condition~\eqref{initial-condition},
we have
\[
\frac{\p^2\log\tau_{\rm BGW}}{\p t_1^2}\left({\bf 0}\right)
=\Omega^{\rm GD}_{1,1}\mid_{{\bf t}={\bf 0}}=\frac{v_1({\bf 0})}{r}=\frac{d_1}r.
\]
The theorem is proved.	
\end{prf}

\section{$W$-constraints of the second kind}\label{section4}

In this section, we show that the generalized BGW tau-function $\tau_{\rm BGW}$ 
for the GD hierarchy satisfies the 
$W$-constraints of the second kind.
Let $v_{\rm BGW}$ be the generalized BGW solution to the GD hierarchy, 
and $u_{\rm BGW}$ the corresponding solution to the KP hierarchy (cf.~\eqref{KP-GD}). 
By Theorem~\ref{equiv-red-DZ} we know that 
$\tau_{\rm BGW}$ can be regarded as a tau-function of~$u_{\rm BGW}$ to the KP hierarchy. 

Let  
\[
L_{\rm KP} = \p + u_{1, {\rm BGW}} \p^{-1} +  u_{2, {\rm BGW}} \p^{-2}  + \cdots  
\]
be the Lax operator for $u_{\rm BGW}$, and let $\Phi$ be the dressing operator for $u_{\rm BGW}$ such that 
$\Phi$ and $\tau_{\rm BGW}$ form a dressing pair. 
The identities in the following lemma are analogues of~\eqref{residue-ml0}. 

\begin{lem}\label{lem-residue-ml}
The following identities hold true: $\forall\,i,k\geq0$, 
\beq\label{hat-ml}
\res_{\p}(M-1)^i \circ L^{\frac{k} r} =\frac{1}{i+1}\p
\Biggl(
\frac{\res_{\lambda} \lambda^k \p_{\mu}^{i+1} \circ  X(\tilde {\mathbf t}_{\rm KP};\lambda,\mu)
\left(\tau_{\rm BGW}\right)\big|_{\mu=\lambda}}{\tau_{\rm BGW}}
\Biggr),
\eeq
where the operator $M$ is given by~\eqref{OS}, and $\tilde {\bf t}_{\rm KP}=\left(t_1-1,t_2,t_3,\dots\right)$.
\end{lem}
\begin{prf} 
Recalling the definition~\eqref{def-X}, 
we have
\beq
\p_\mu^{i+1} \circ X\bigl(\tilde {\bf t}_{\rm KP};\lambda,\mu\bigr)
=\sum_{j=0}^{i+1}\binom{i+1}j(-1)^{i+1-j}
\p_\mu^j \circ X({\bf t}_{\rm KP};\lambda,\mu).
\eeq
By using~\eqref{residue-ml0}, 
one can then write the right-hand side of~\eqref{hat-ml} into 
\beq
\frac1{i+1}\sum_{j=1}^{i+1}\binom{i+1}j (-1)^{i+1-j}j M^{j-1} \circ L_{\rm KP}^k
=\res_{\p}(M-1)^i \circ L_{\rm KP}^k.
\eeq
By noticing that $L_{\rm KP}=L^{\frac1r}$, the lemma is proved.
\end{prf}

\begin{lem}\label{minus-zero-2}
The following identities hold true:
\beq\label{w-cons-0}
\left((M-1)^{i} \circ L^{k+\frac {i} r}\right)_-=0,
\quad
\forall \, i,k\geq0.
\eeq
\end{lem}
\begin{prf}
The case $i=0$ is obvious.
For $i\geq1$,
let us first show $\bigl((M-1) L^{1/r}\bigr)_-=0$. 
Observe that 
\beq
\sum_{i\geq1}i t_i\frac{\p}{\p t_i}-\frac{\p}{\p t_1}+\frac{d_1}{r}
=\res_{\lambda}\biggl(\frac{\lambda}{2}\p_{\mu}^2X({\bf t}_{\rm KP};\lambda,\mu)
- \lambda \p_{\mu}X({\bf t}_{\rm KP};\lambda,\mu)
+\frac{d_1}{\lambda r}\biggr) \bigg|_{\mu=\lambda} d\lambda
=:G({\bf t}_{\rm KP}).
\eeq
By identity~\eqref{new-string}, we have
$G({\bf t}_{\rm KP})(\tau_{\rm BGW})=0$.
Therefore,
\begin{align}
0=&\res_\nu \frac{X({\bf t}_{\rm KP};\nu)\circ
	G({\bf t}_{\rm KP})(\tau_{\rm BGW}({\bf t}_{\rm KP}))}{\tau_{\rm BGW}({\bf t}_{\rm KP})}
\frac{ X^*({\bf t}_{\rm KP}'; \nu)\left(\tau_{\rm BGW}({\bf t}_{\rm KP}')\right)}{\tau_{\rm BGW}({\bf t}_{\rm KP}')} d \nu \nn\\
=&-\res_{\lambda}\left(\lambda \p_{\lambda}(\psi({\bf t}_{\rm KP};\lambda))\psi^*({\bf t}_{\rm KP}';\lambda)
-\lambda \psi({\bf t}_{\rm KP};\lambda)\psi^*({\bf t}_{\rm KP}';\lambda)\right) d\lambda.\label{negative-1-1}
\end{align}
Here the second equality used the identity~\eqref{double-wave}. 
By using the facts 
\[
M(\psi)=\p_\lambda (\psi),
\quad
L_{\rm KP}(\psi)=\lambda \psi,
\]
as well as the following identity~\cite{DJKM}: for arbitrary $x,x'$ and arbitrary $t_2=t_2', t_3=t_3',\dots$,
\begin{align}
&\left(U({\bf t}_{\rm KP}) V({\bf t}_{\rm KP}')\right)_- \left(\delta(x-x') \right)\nn\\
&\quad =-\res_z U({\bf t}_{\rm KP})\left(e^{xz+\sum_{i\geq2}t_i z^i}\right) V^*({\bf t}_{\rm KP}')\left(e^{-x'z-\sum_{i\geq2}t_i z^i} \right)dz\, H(x-x')
\end{align}
with $U({\bf t}_{\rm KP}),V({\bf t}_{\rm KP}')$ being arbitrary pseudo-differential operators whose coefficients are 
power series of their arguments, we have
\begin{align}
0&=-\res_{\lambda}\left(\lambda \p_{\lambda}(\psi({\bf t}_{\rm KP};\lambda)) \psi^*({\bf t}_{\rm KP}';\lambda)
-\lambda \psi({\bf t}_{\rm KP};\lambda)\psi^*({\bf t}_{\rm KP}';\lambda)\right) d\lambda\,H(x-x') \nn\\
& =\left(M \circ L_{\rm KP}-L_{\rm KP}\right)_-\left(\delta(x-x')\right) \nn \\
& =\sum_{i\geq1} {\rm Coef}_{\p^{-i}}\Big(M \circ L_{\rm KP}-L_{\rm KP}\Big) \frac{(x-x')^{i-1}}{(i-1)!}H(x-x').
\end{align}
Here $\delta(x)$ is the Dirac delta function, and $H(x)$ is the Heaviside unit step function. 
Therefore, $\bigl((M-1) \circ L_{\rm KP}\bigr)_-= 0$.

 By using the fact that
\[
 (M-1)^{i+1} \circ L^{k+\frac{i+1}r}
= M^i \circ L^{k +\frac i r}\circ (M-1)\circ L
-\left(k r+ i\right) (M-1)^i\circ L^{k+\frac i r},\quad \forall\,i,k\geq0,
\]
we can prove by induction that identities~\eqref{w-cons-0} are true.
\end{prf}

From Lemmas~\ref{lem-residue-ml} and \ref{minus-zero-2}, 
it follows that, for $\alpha=1,2,\dots,r-1$ and $q\geq\alpha$,
\beq\label{constraint-0}
\p_x\left(
\frac{
\res_{\lambda}
\lambda^{\alpha+ (q-\alpha) r}
\Bigl(\p_{\mu}^{\alpha+1} \circ X\bigl(\tilde {\bf t}_{\rm KP};\lambda,\mu\bigr)\Bigr)\left(\tau_{\rm BGW}({\bf t})\right)\Big|_{\mu=\lambda}}{\tau_{\rm BGW}({\bf t})}\right)=0.
\eeq
Denote
\beq\label{vertex-GD}
X_{\rm GD}({\bf t};\lambda,\mu):=
e^{\sum_{i\in\NN\backslash r\NN}  t_i\left(\mu^i-\lambda^i\right)}
e^{\sum_{i\in\NN\backslash r\NN}\left(\frac1{i\lambda^i}-\frac1{i\mu^i}\right)\frac{\p}{\p t_i}}. 
\eeq
We have that
\begin{align*}
&\p_{\mu}^{\alpha+1}
\circ
X_{\rm GD}( {\bf t};\lambda,\mu)
\circ
e^{\sum_{j\geq1}\frac1{jr}\left(\frac{1}{\lambda^{jr}}-\frac{1}{\mu^{jr}}\right)
	\frac{\p}{\p t_{jr}}}\\
=&
\p_{\mu}^{\alpha+1}
\circ
e^{-\sum_{i\geq1}t_{ir}(\mu^{ir}-\lambda^{ir})}
\circ
X( {\bf t}_{\rm KP};\lambda,\mu)
\\
=&\sum_{k_1=0}^{\alpha+1}\binom{\alpha+1}{k_1}
\p_{\mu}^{k_1} 
\left(e^{-\sum_{j\geq1}t_{jr}(\mu^{jr}-\lambda^{jr})}\right)
\p_{\mu}^{\alpha+1-{k_1}} \circ X( {\bf t}_{\rm KP};\lambda,\mu)
\\
=&\sum_{k_1=0}^{\alpha+1}\sum_{p\geq1}f_{k_1,p}\mu^{pr-k_1}
\p_{\mu}^{\alpha+1-k_1} \circ  X( {\bf t}_{\rm KP};\lambda,\mu),
\end{align*}
where $f_{k_1,p}=f_{k_1,p}(t_{r},t_{2r},\dots)\in\CC[[t_r,t_{2r},\dots]]$. 
By using this identity, one can obtain that
\begin{align*}
&\frac{1}{\tau_{\rm BGW}({\bf t})}\res_{\lambda=\infty}\lambda^{\alpha+(q-\alpha)r}
\Big(\p_{\mu}^{\alpha+1} \circ X_{\rm GD}\bigl(\tilde {\bf t};\lambda,\mu\bigr)\big(\tau_{\rm BGW}({\bf t})\big)\Big)\Big|_{\mu=\lambda} d\lambda \nn \\
=&\frac1{\tau_{\rm BGW}({\bf t})}\sum_{j=0}^{\alpha+1}\sum_{p\geq1}f_{j,p} \res_{\lambda=\infty}
\lambda^{\alpha-j+(q-\alpha+p)r}\Big(\p_\mu^{\alpha-j+1}\circ  X\bigl(\tilde {\bf t}_{\rm KP};\lambda,\mu\bigr)\big(\tau_{\rm BGW}({\bf t})\big)\Big)\Big|_{\mu=\lambda}d\lambda.
\end{align*}
Together with~\eqref{constraint-0}, it follows that
\beq\label{constraint-1}
\p_x\left(\frac{W^{\rm red}_{\alpha,q}({\bf t})\left(\tau_{\rm BGW}({\bf t})\right)}{\tau_{\rm BGW}({\bf t})}\right)=0,\qquad 
\alpha=1,\dots,r-1,\quad q\geq \alpha,
\eeq
where the operators $W^{\rm red}_{\alpha,q}({\bf t})$ are defined in~\eqref{defJ}.

We are ready to prove Theorem~\ref{thmwconst}.

\begin{prfn}{Theorem~\ref{thmwconst}}
From formula~\eqref{constraint-1},
we know that 
there exist power series $c_{\alpha,q}({\bf t})$, independent of $x=t_1$, such that
\beq\label{w-f}
W^{\rm red}_{\alpha,q}({\bf t})\left(\tau_{\rm BGW}({\bf t})\right)
=c_{\alpha,q}({\bf t})\tau_{\rm BGW}({\bf t}).
\eeq
By using Theorem~\ref{exis-bgw-tau} 
and the fact that
$W^{\rm red}_{1,1}({\bf t})=\sum_{i\in\NN\backslash r\NN}i\tilde t_i\frac{\p}{\p t_i}$,
we have $\rho_1=d_1/r$.
By the definition~\eqref{defJ},
one can verify that
\[
\biggl[W^{\rm red}_{1,1}({\bf t}), W^{\rm red}_{\alpha,q}({\bf t})\biggr]=-(q-\alpha) r W^{\rm red}_{\alpha,q}({\bf t}).
\]
Applying the both sides of this identity onto~$\tau_{\rm BGW}({\bf t})$,
we obtain that
\begin{align*}
W^{\rm red}_{1,1}({\bf t})\left(c_{\alpha,q}({\bf t})\right)=-(q-\alpha) r c_{\alpha,q}({\bf t}). 
\end{align*}
This implies that $c_{\alpha,q}({\bf t})$ are power series with non-positive degrees 
if we let the degree of~$t_i$ be assigned with~$i$, $i\in\ZZ_{\geq2}\backslash r\NN$. 
So~$c_{\alpha,q}({\bf t})$ must be constants,
moreover, these constants vanish if~$q>\alpha$.

Let us proceed to prove the property~\eqref{prprho}.
To this end, we first prove the following lemma.
\begin{lem}\label{corW}
Denote
\beq\label{def-W}
S_{\alpha,q}
:=\frac1{\alpha+1}\res_{\lambda}\lambda^{\alpha+(q-\alpha)r}: \Biggl(\sum_{j\in\NN\backslash r\NN}j\lambda^{j-1}
\tilde t_j
+\sum_{j\in\NN\backslash r\NN}\lambda^{-j-1}\frac{\p}{\p t_j}
\Biggr)^{\alpha+1}: \, ,
\eeq
where $\alpha=1,\dots,n$, $q\geq0$, 
and ``$:\;:$" denotes the normal ordering (defined by putting the operators $\frac{\p}{\p t_j}$ on the right of operators $\tilde t_i$). 
The constraints~\eqref{w-eq-1} can be equivalently written as
\beq\label{weq2}
S_{\alpha,q} \bigl(\tau_{\rm BGW}\bigr)=(-1)^\alpha \sigma_\alpha \delta_{\alpha,q}\tau_{\rm BGW}, \qquad \alpha=1,\dots,r-1,~ q\geq \alpha,
\eeq
where $\sigma_1,\dots,\sigma_{r-1}$ are certain polynomials of $\rho_1,\dots,\rho_{r-1}$.
\end{lem}
\begin{prf}
We denote
\[
a({\bf t};\lambda):=\sum_{i\in\NN\backslash r\NN}\lambda^i t_i,\quad
b({\bf t};\lambda):=-\sum_{i\in\NN\backslash r\NN}\frac{1}{i\lambda^i}\frac{\p}{\p t_i},
\]
and denote
\beq
 P_i({\bf t};\lambda)
 :=\p_{\mu}^{i+1}\Bigl(
 e^{a({\bf t};\mu)-a({\bf t};\lambda)}
 e^{b({\bf t};\mu)-b({\bf t};\lambda)}\Bigr)\bigg|_{\mu=\lambda},
 \quad
 i\geq0.
\eeq
It is easy to see that $P_i({\bf t};\lambda)$ satisfy following the recursion relations:
\beq\label{relationP}
P_i({\bf t};\lambda)
=\frac{\p a({\bf t};\lambda)}{\p \lambda} \circ P_{i-1}({\bf t};\lambda)+P_{i-1}({\bf t};\lambda) \circ \frac{\p b({\bf t};\lambda)}{\p\lambda}
+\frac{\p P_{i-1}({\bf t};\lambda)}{\p \lambda}.
\eeq
By using the above~\eqref{relationP}, one can prove the following identity by induction:
\beq
P_i({\bf t};\lambda)\;=\; \sum_{j=0}^{i-1}\frac{\p^j}{\p\lambda^j} \biggl(:\bigl(\p_{\lambda}(a({\bf t};\lambda))+\p_{\lambda}(b({\bf t};\lambda))\bigr)^{i+1-j}:\biggr),
\quad
i\geq1.
\eeq
Then we have that, for $\alpha=1,\dots,r-1$ and $q\geq \alpha$, 
\begin{align*}
&\res_{\lambda}\lambda^{\alpha+(q-\alpha)r}
:\left(\p_{\lambda}a({\bf t};\lambda)+\p_{\lambda}b({\bf t};\lambda)\right)^{i+1}:d\lambda\nn\\
=&\res_{\lambda}\lambda^{\alpha+(q-\alpha)r}P_\alpha({\bf t};\lambda)
-\sum_{j=1}^{\alpha-1}\res_{\lambda}\lambda^{\alpha+(q-\alpha)r}
:\p_{\lambda}^j\left(\p_{\lambda}a({\bf t};\lambda)+\p_{\lambda}b({\bf t};\lambda)\right)^{\alpha+1-j}: d\lambda \nn\\
=&\res_{\lambda}\lambda^{\alpha+(q-\alpha)r}P_\alpha({\bf t};\lambda) d\lambda\\
&-\sum_{j=1}^{\alpha-1}
\frac{(-1)^j(\alpha+(q-\alpha)r)!}{(\alpha+(q-\alpha)r-j)!}
\res_{\lambda}\lambda^{\alpha-j+(q-\alpha)r}
:\left(\p_{\lambda}a({\bf t};\lambda)+\p_{\lambda}b({\bf t};\lambda)\right)^{\alpha+1-j}: d\lambda.
\end{align*}
By using the definition~\eqref{def-W} and by noticing that $W^{\rm red}_{\alpha,q}({\bf t})$ can be rewritten as
\beq
W^{\rm red}_{\alpha,q}({\bf t})=
\frac{1}{\alpha+1}
\res_{\lambda}\lambda^{\alpha+(q-\alpha)r}P_\alpha({\bf t};\lambda),
\eeq
we have 
\beq
S_{\alpha,q}=W^{\rm red}_{\alpha,q}-\sum_{j=1}^{\alpha-1}
\frac{(-1)^j(\alpha-j+1)(\alpha+(q-\alpha)r)!}{(\alpha+(k-\alpha)r-j)!} S_{\alpha-j,q-j}.
\eeq
Therefore, by using Theorem~\ref{thmwconst} we obtain~\eqref{w-eq-1}, where the constants 
$\sigma_1,\dots,\sigma_{r-1}$ can be uniquely determined by
\beq\label{rhosigma}
\sigma_\alpha
= \rho_\alpha+\frac{\alpha!}{\alpha+1}\sum_{j=1}^{\alpha-1}
\frac{(-1)^j}{j!}\sigma_j,
\quad
\alpha=1,\dots,r-1.
\eeq
The lemma is proved.
\end{prf}

The following lemma will also be needed,
and will also have other important applications.
For simplicity, we denote 
\[\langle\tau_{i_1}\cdots\tau_{i_k}\rangle^\bullet:= \left.\frac{\p^k \tau_{\rm BGW}}{\p t_{i_1} \dots \p t_{i_k}} \right|_{\bt ={\bf 0}}, \quad 
\langle\tau_{i_1}\cdots\tau_{i_k}\rangle:= \left.\frac{\p^k \log\tau_{\rm BGW}}{\p t_{i_1} \dots \p t_{i_k}} \right|_{\bt = {\bf 0}}. \]

\begin{lem}\label{uniqueness}
The system~\eqref{w-eq-1} (or equivalently~\eqref{weq2}) has a unique solution in $\CC[[{\bf t}]]$ with initial value~$1$. 
\end{lem}
\begin{prf}
The existence of the solution is already proved.
To show the uniqueness, we use the argument similar to that in~\cite{AM, BBCCN, LYZ}.
By~\eqref{def-W}, we know that 
\begin{align*}
S_{\alpha,q}
=&\sum_{j=0}^\alpha\frac{\alpha!}{j!(\alpha+1-j)!}
\sum_{k_{j+1}+\cdots+k_{\alpha+1}\atop-k_1-\cdots-k_j=(q-\alpha)r} 
k_1\cdots k_j  \tilde t_{k_1}\cdots  \tilde t_{k_j} \frac{\p^{\alpha-j}}{\p t_{k_{j+1}}\cdots \p t_{k_{\alpha+1}}}\\
=&\sum_{j=0}^\alpha (-1)^j\binom \alpha j
\sum_{p=0}^{\alpha-j}\sum_{k_{p+1}+\cdots+k_{\alpha-j+1}\atop-k_1-\cdots-k_p=(q-\alpha)r+j} 
k_1\cdots k_p t_{k_1}\cdots t_{k_p} \frac{\p^{\alpha-j-p}}{\p t_{k_{p+1}}\cdots \p t_{k_{\alpha-j+1}}},
\end{align*}
where $\alpha=1,\dots,r-1$ and $q\geq\alpha$.
Hence equations~\eqref{weq2} can be recast to
\begin{align}
\frac{\p\tau_{\rm BGW}}{\p t_{\alpha+(q-\alpha)r}}
=&\sigma_\alpha\delta_{q,\alpha}\tau_{\rm BGW} + \sum_{j=0}^\alpha (-1)^{\alpha-j}\binom \alpha j
\sum_{p=0}^{\alpha-j}\sum_{k_{p+1}+\cdots+k_{\alpha-j+1}\atop-k_1-\cdots-k_p=(q-\alpha)r+j} \frac{\p^{\alpha-j-p} \tau_{\rm BGW}}{\p t_{k_{p+1}}\cdots \p t_{k_{\alpha-j+1}}}  \prod_{a=1}^p 
k_a t_{k_a} .
\end{align}
In terms of $\langle\tau_{i_1}\cdots\tau_{i_k}\rangle^\bullet$,
we have the recursion relations:
\begin{align}
&\langle\tau_A\tau_{\alpha+(m-\alpha)r}\rangle^\bullet \nn\\
=&\langle\tau_A\rangle^\bullet+\sum_{j=0}^\alpha (-1)^{\alpha-j} \binom \alpha j
\sum_{p=0}^{\alpha-j}\sum_{k_{p+1}+\cdots+k_{(\alpha-j)+1}\atop-k_1-\cdots-k_p=(m-\alpha)r+j}
p!k_1\cdots k_p \langle\tau_{A\backslash \{k_1,\dots,k_p\}\cup \{k_{p+1},\dots,k_{\alpha-j+1}\}}
\rangle^\bullet.
\label{correlator-recursion}
\end{align}
Here $\tau_A:=\tau_{a_1}\cdots \tau_{a_N}$ for $A=\{a_1,\dots,a_N\}$. 
In this way it is clear that all the coefficients of $\langle\tau_{i_1}\cdots\tau_{i_k}\rangle^\bullet$ can be uniquely determined. 
The lemma is proved. 
\end{prf}

(The uniqueness statement in the above Lemma~\ref{uniqueness} can also be proved directly from~\eqref{w-eq-1}.)

By using Lemma~\ref{corW} and formula~\eqref{correlator-recursion}, we obtain that 
\beq
\langle\tau_1\tau_\alpha\rangle
=\langle\tau_\alpha\rangle
=c_\alpha(\sigma_1,\dots,\sigma_\alpha)= \sigma_\alpha+\gamma_\alpha\left(\sigma_1,\dots,\sigma_{\alpha-1}\right),
\eeq
where $c_\alpha$ are certain polynomials of $\sigma_1,\dots,\sigma_\alpha$. 
The property~\eqref{prprho} then follows from~\eqref{rhosigma} and the relation
\beq
\frac{\p\log\tau_{\rm BGW}}{\p t_1\p t_\alpha}=\Omega_{1,\alpha}(v_{\rm BGW},v_{x,\rm BGW},\dots)=\res_{\p} L^{\frac{\alpha}{r}}.
\eeq
The theorem is proved.
\end{prfn}

\smallskip 

We note that, the constants $c_\alpha$ in the above proof are initial values of 
the {\it normal coordinates} $\Omega_{1,\alpha}=\res_{\p} L^{\alpha/r}$
for the GD hierarchy, i.e., of the corresponding Dubrovin--Zhang hierarchy~\cite{DZ-norm} (cf.~also~\cite{BY}).

\smallskip

By using Theorem~\ref{thmwconst}, Lemma~\ref{corW} and~\ref{uniqueness}, we arrive at the following theorem. 

\begin{thm}\label{thm-main}
A power series $\tau\in\CC[[{\mathbf t}]]$ satisfies~\eqref{w-eq-1} 
if and only if 
$\tau$ is the tau-function for the Gelfand--Dickey hierarchy satisfying~\eqref{new-string}.
\end{thm}

\section{Examples} \label{section5}
In this section,
we use Theorem~\ref{thmwconst} (in particular Lemma~\ref{corW}) to compute
$\tau_{\rm BGW}$ and $\log\tau_{\rm BGW}$. 
\begin{emp}
	For the case with $r=2$, the constraints~\eqref{weq2} give the following relations:
	\begin{align*}
	\langle\prod_{i=1}^N 
	\tau_{2a_i+1}\tau_{2m+1}\rangle^\bullet
	=&\sum_{i=1}^N (2a_i+1)\langle\tau_{2a_i+2m+1}\prod_{j\neq i}\tau_{2a_j+1}\rangle^\bullet
	+\frac12\sum_{k_1+k_2=m-1}\langle\tau_{2k_1+1}\tau_{2k_2+1}\prod_{i=1}^N\tau_{2a_i+1}\rangle^\bullet\\
	&+\delta_{m,0}c_1\langle\prod_{i=1}^N\tau_{2a_i+1}\rangle^\bullet,
	\end{align*}
	where $N, a_1,\dots,a_N, m\geq0$. 
	The constants $c_1,d_1, \sigma_1,\rho_1$ are related by 
\beq
\rho_1=\sigma_1=c_1,\quad d_1=2c_1.
\eeq
We have
\begin{align*}
&\langle\tau_1\rangle^\bullet=c_1,
\quad
\langle\tau_1^2\rangle^\bullet=c_1(c_1+1),
\quad
\langle\tau_3\rangle^\bullet=\frac12c_1(c_1+1), \quad \langle\tau_1^3\rangle^\bullet
=c_1(c_1+1)(c_1+2),\\
&\langle\tau_1\rangle=c_1,
\quad
\langle\tau_1^2\rangle=c_1,
\quad
\langle\tau_3\rangle=\frac12c_1(c_1+1), \quad \langle\tau_1^3\rangle
=2c_1,
\quad
\langle\tau_3\tau_1\rangle
=\frac32c_1(c_1+1).
\end{align*}
\end{emp}
	
\begin{emp}
For the case with $r=3$, 
the constants $c_\alpha,d_\alpha, \sigma_\alpha,\rho_\alpha$ are related by 
	\begin{align*}
	&\rho_1=\sigma_1= c_1,\quad d_1=3c_1,\\
	&\rho_2=c_2+\frac23c_1,\quad \sigma_2= c_2,\quad d_2=\frac32 c_2+3c_1.\\
	\end{align*}
We have
\begin{align*}
&\langle\tau_1\rangle^\bullet
=c_1,
\quad
\langle\tau_2\rangle^\bullet
=c_2,
\quad
\langle\tau_1^2\rangle^\bullet
=c_1(c_1+1),\\
&\langle\tau_2\tau_1\rangle^\bullet
=(c_1+2)\frac{c_2}2,
\quad
\langle\tau_1^3\rangle^\bullet
=c_1(c_1+1)(c_1+2),\\
&\langle\tau_4\rangle^\bullet
=c_2(c_1+2),
\quad
\langle\tau_2^2\rangle^\bullet
=c_2^2-2c_1(c_1+1),\\
&\langle\tau_2\tau_1^2\rangle^\bullet
=c_2(c_1+2)(c_1+3),
\quad 
\langle\tau_1^4\rangle^\bullet
=c_1(c_1+1)(c_1+2)(c_1+3),
\\
&\langle\tau_1\rangle
=c_1,
\quad
\langle\tau_2\rangle
=c_2,
\quad
\langle\tau_1^2\rangle
=c_1,\quad
\langle\tau_2\tau_1\rangle
=2c_2,
\quad
\langle\tau_1^3\rangle
=2c_1,\\
&\langle\tau_4\rangle
=c_2(c_1+2),
\quad
\langle\tau_2^2\rangle
=-2c_1(c_1+1), \quad 
\langle\tau_2\tau_1^2\rangle
=6c_2,
\quad 
\langle\tau_1^4\rangle
=6c_1.
\end{align*}
\end{emp}

\begin{emp}
Similarly, for the case with $r=4$, we have
\begin{align*}
&\rho_1=\sigma_1=c_1,\quad 
d_1=4c_1, \\
&\rho_2=c_2+\frac23 c_1,\quad 
\sigma_2=c_2\quad 
d_2=4c_2+8 c_1, \\
&\rho_3=c_3-\frac34 c_2-\frac32 c_1^2,\quad 
\sigma_3=c_3-\frac32 c_1^2-\frac32 c_1\quad 
d_3=\frac43c_3+3c_2+2c_1^2+10c_1,
\end{align*}
and
\begin{align*}
&\langle\tau_1\rangle^\bullet
=c_1,
\quad
\langle\tau_2\rangle^\bullet
=c_2,
\quad
\langle\tau_1^2\rangle^\bullet
=c_1(c_1+1),
\quad
\langle\tau_3\rangle^\bullet
=c_3,
\\&
\langle\tau_2\tau_1\rangle^\bullet
=c_2(c_1+2), 
\quad
\langle\tau_1^3\rangle^\bullet
=c_1(c_1+1)(c_1+2),
\\&
\langle\tau_3\tau_1\rangle^\bullet
=c_3(c_1+3),
\quad
\langle\tau_2^2\rangle^\bullet
=4c_3+c_2^2-2c_1(c_1+1),
\\&
\langle\tau_2\tau_1^2\rangle^\bullet
=c_2(c_1+2)(c_1+3),
\quad
\langle\tau_1^4\rangle^\bullet
=c_1(c_1+1)(c_1+2)(c_1+3),
\\
&\langle\tau_1\rangle
=c_1,
\quad
\langle\tau_2\rangle
=c_2,
\quad
\langle\tau_1^2\rangle
=c_1, \quad 
\langle\tau_3\rangle
=c_3,
\quad
\langle\tau_2\tau_1\rangle
=2c_2, \quad 
\langle\tau_1^3\rangle
=2c_1,\\
& \langle\tau_3\tau_1\rangle
=3c_3, \quad 
\langle\tau_2^2\rangle
=4c_3-2c_1(c_1+1),
\quad
\langle\tau_2\tau_1^2\rangle
=6c_2,
\quad
\langle\tau_1^4\rangle
=6c_1.
\end{align*}
\end{emp}

\begin{rmk}
Define 
\beq
\langle\tau_{i_1}\cdots\tau_{i_k}\rangle_{\infty} = \lim_{r\rightarrow \infty} \langle\tau_{i_1}\cdots\tau_{i_k}\rangle. 
\eeq
We call $\langle\tau_{i_1}\cdots\tau_{i_k}\rangle_{\infty}$ the stabilized generalized BGW correlators. 
For example,
\begin{align*}
&\langle\tau_1\rangle_{\infty}
=c_1,
\quad
\langle\tau_2\rangle_{\infty}
=c_2,
\quad
\langle\tau_1^2\rangle
=c_1,
\quad
\langle\tau_3\rangle_{\infty}
=c_3,
\\&
\langle\tau_2\tau_1\rangle_{\infty}
=2c_2,
\quad
\langle\tau_1^3\rangle_{\infty}
=2c_1,
\quad
\langle\tau_4\rangle_{\infty}
=c_4,
\quad
\langle\tau_3\tau_1\rangle_{\infty}
=3c_3,
\\&
\langle\tau_2^2\rangle_{\infty}
=4c_3-2c_1(c_1+1),
\quad
\langle\tau_2\tau_1^2\rangle_{\infty}
=6c_2,
\quad
\langle\tau_1^4\rangle_{\infty}
=6c_1.
\end{align*}
The partition function of these stabilized correlators and its relation to the KP hierarchy deserves a further study.
\end{rmk}

\medskip

\noindent D.Y.: School of Mathematical Sciences, USTC, Hefei 230026, P.R. China

\noindent diyang@ustc.edu.cn

\smallskip

\noindent C.Z.: Institute of Geometry and Physics, USTC, Hefei 230026, P.R. China

\noindent zhouch@ustc.edu.cn


\begin{thebibliography}{99}
	
\bibitem{ASM}
Adler, M., Shiota, T., van Moerbeke, P., 
A Lax representation for the vertex operator and the central extension. Comm. Math. Phys. {\bf 171} (1995), 547--588.

\bibitem{AM}
Adler, M., van Moerbeke, P., A matrix integral solution to two-dimensional $W_p$-gravity, 
Comm. Math. Phys. {\bf 147} (1992), 25--56.

\bibitem{A}
Alexandrov, A., 
Cut-and-join description of generalized Brezin-Gross-Witten model. Adv. Theor. Math. Phys.~{\bf 22} (2018), 1347--1399. 

\bibitem{BM0}  
Bakalov, B., Milanov, T., $\mathcal{W}_{N+1}$-constraints for singularities of type $A_N$, arXiv:0811.1965.

\bibitem{BM} 
Bakalov, B., Milanov, T., 
$\mathcal {W} $-constraints for the total descendant potential of a simple singularity. 
Compos. Math. {\bf 149} (2013), 840--888.

\bibitem{BDY}
Bertola, M., Dubrovin, B., Yang, D., 
Simple Lie algebras, Drinfeld--Sokolov hierarchies, and multi-point correlation functions, Mosc. Math. J.~{\bf 21} (2021), 233--270.

\bibitem{BR}
Bertola, M., Ruzza, G., Brezin-Gross-Witten tau function and isomonodromic deformations. 
Commun. Number Theory Phys.~{\bf 13} (2019), 827--883.

\bibitem{BY}
Bertola, M., Yang, D., The partition function of the extended $r$-reduced Kadomtsev-Petviashvili hierarchy.
J. Phys. A~{\bf 48} (2015), 195205, 20 pp. 

\bibitem{BBCCN}
Borot, G., Bouchard, V., Creutzig, T., Chidambaram, N. K., Noshchenko, D., 
Higher Airy structures, $\mathcal{W}$ algebras and topological recursion. 
arXiv:1812.08738.

\bibitem{BG}
Brezin, E., Gross, D. J.,
The external field problem in the large N limit of QCD. Phys. Lett. B~{\bf 97} (1980), 120--124.

\bibitem{Buryak}
Buryak, A.,
 Open intersection numbers and the wave function of the KdV hierarchy. Mosc. Math. J.~{\bf 16} (2016), 27--44.

\bibitem{DJKM}
Date, E., Jimbo, M., Kashiwara, M., Miwa, T., 
Transformation groups for soliton equations, 
in Non-linear integrable systems--Classical theory and quantum theory, World Sci., Singapore, 1983.

\bibitem{DJM}
Date, E., Jimbo, M., Miwa, T., 
Solitons: Differential equations, symmetries and infinite dimensional algebras, 
Cambridge: Cambridge University Press, 2000.

\bibitem{Dickey}
Dickey, L. A.,
Soliton equations and Hamiltonian systems, 2nd ed, World Sci., 2003.

\bibitem{DN}
Do, N., Norbury, P., 
Topological recursion on the Bessel curve. Commun. Number Theory Phys.~{\bf 12} (2018), 53--73. 

\bibitem{DS}
Drinfeld, V. G., Sokolov, V. V.,
Lie algebras and equations of Korteweg--de Vries type, 
J. Math. Sci. {\bf 30}, (1985), 1975--2036.
Translated from Itogi Nauki i Tekhniki, Seriya Sovremennye Problemy Matematiki (Noveishie Dostizheniya) {\bf 24} (1984), 81--180.

\bibitem{Du}
Dubrovin, B., 
Geometry of 2D topological field theories. Integrable systems and quantum groups (Montecatini Terme, 1993), 120--348, 
Lecture Notes in Math. 1620, Springer, Berlin, 1996. 

\bibitem{DYZ}
Dubrovin, B., Yang, D., Zagier, D., 
On tau-functions for the KdV hierarchy. Selecta Math.~{\bf 27} (2021), Paper No. 12, 47~pp.

\bibitem{DZ-norm}
Dubrovin, B., Zhang, Y., Normal forms of hierarchies of integrable PDEs, Frobenius manifolds
and Gromov-Witten invariants. arXiv:math/0108160.

\bibitem{DG}
Duistermaat, J.~J., Gr\"unbaum, F.~A., Differential equations in the spectral parameter. Comm. Math. Phys.,
{\bf 103} (1986), 177--240.

\bibitem{FF} 
Feigin, B., Frenkel, E., Integrals of motion and quantum groups. 
Integrable systems and quantum groups (Montecatini Terme, 1993), 349--418,
Lecture Notes in Math. 1620, Springer, Berlin, 1996.

\bibitem{FI} 
Frenkel, I.~B., Representations of affine Lie algebras, Hecke modular
forms and Korteweg-de Vries type equations. Lie algebras and related topics (New
Brunswick, N.J., 1981), pp. 71--110,  Lecture Notes in Math., 933, Springer,
Berlin-New York (1982).

\bibitem{FSZ} 
Faber, C., Shadrin, S., Zvonkine, D., Tautological relations and the $r$-spin Witten conjecture.
Ann. Sci. \'Ec. Norm. Sup\'er. {\bf 43} (2010), 621--658.

\bibitem{FL} 
Fateev, V. A., Lukyanov, S. L., The models of two-dimensional conformal quantum field theory with $Z_n$ symmetry.
Internat. J. Modern Phys. A {\bf 3} (1988), 507--520.

\bibitem{FKN}
Fukuma, M., Kawai, H., Nakayama, R., 
Infinite-dimensional Grassmannian structure of two-dimensional quantum gravity. Comm. Math. Phys.~{\bf 143} (1992), 371--403.

\bibitem{G}
Givental, Alexander $A_{n-1}$ singularities and $n$KdV hierarchies. 
Dedicated to Vladimir I. Arnold on the occasion of his 65th birthday. Mosc. Math. J.~{\bf 3} (2003), 475--505.

\bibitem{GM}
Givental, A., Milanov, T., Simple singularities and integrable hierarchies. 
The breadth of symplectic and Poisson geometry, 173--201, Progr. Math.,~{\bf 232}, Birkh\"auser Boston, Boston, MA, 2005. 
 
\bibitem{Goe}
Goeree, J., W-constraints in 2D quantum gravity, Nuclear Phys. B~{\bf 358} (1991), 737--757.

\bibitem{GN}
Gross, D. J., Newman, M. J., 
Unitary and hermitian matrices in an external field. II. The Kontsevich model and continuum Virasoro constraints. Nuclear Phys. B~{\bf 380} (1992), 
168--180.

\bibitem{GW}
Gross, D. J.,  Witten, E., 
Possible Third Order Phase Transition in the Large N Lattice Gauge Theory. Phys. Rev. D~{\bf 21} (1980), 446--453.

\bibitem{Kac} 
Kac, V., Vertex algebras for beginners. Second edition. University Lecture Series, 10.
American Mathematical Society, Providence, RI, 1998.
 
\bibitem{KJM}
Kashiwara, M., Jimbo, M., Miwa, T., Transformation groups for soliton equations. Bosonization, 1982.

\bibitem{KN}
Kazarian, M., Norbury, P., 
Polynomial relations among kappa classes on the moduli space of curves. arXiv:2112.11672.

\bibitem{K}
Kontsevich, M., 
Intersection theory on the moduli space of curves and the matrix Airy function, Comm. Math. Phys.~{\bf 147} (1992), 1--23.

\bibitem{LWZ}
Liu S.-Q., Wu C.-Z., Zhang Y, Virasoro Constraints for Drinfeld-Sokolov hierarchies and equations of Painlev\'e type,
arXiv:1908.06707.

\bibitem{LYZ}
Liu, S.-Q., Yang, D., Zhang, Y., Uniqueness Theorem of $\mathcal{W}$-Constraints for Simple Singularities. 
Lett. Math. Phys. {\bf 103} (2013), 1329--1345.

\bibitem{MMS}
Mironov, A., Morozov, A., Semenoff, G.~W., 
Unitary matrix integrals in the framework of the generalized Kontsevich model. Internat. J. Modern Phys. A~{\bf 11} (1996), 5031--5080.

\bibitem{N}
Norbury, P., A new cohomology class on the moduli space of curves. arXiv:1712.03662.

\bibitem{OS}
Orlov, A. Y., Schulman, E. I., Additional symmetries for integrable an conformal algebra representation. 
Lett. Math. Phys. {\bf 12} (1986), 171--179.

\bibitem{W1}
Witten, E., Two-dimensional gravity and intersection theory on moduli space. Surveys in differential geometry (Cambridge, MA, 1990), 243--310, Lehigh Univ., Bethlehem, PA, 1991.

\bibitem{W2}
Witten, E., Algebraic geometry associated with matrix models of two-dimensional gravity, in Topological methods in modern mathematics 
(Stony Brook, NY, 1991), Publish or Perish, 1993, 235--269.

\bibitem{YZ21}
Yang, D., Zhang, Q., On the Hodge-BGW correspondence. arXiv:2112.12736.

\bibitem{YZ}
Yang, D., Zhou, C., On an extension of the generalized BGW tau-function. Lett. Math. Phys.~{\bf 111} (2021), Paper No.~123, 23~pp.

\bibitem{Zhou}
Zhou, J., Solution of W-constraints for R-spin intersection numbers. arXiv:1305.6991.

\end{thebibliography}
\end{document}